# Design and Implementation of mmWave Surface Wave Enabled Fluid Antennas and Experimental Results for Fluid Antenna Multiple Access

Yuanjun Shen, *Member, IEEE*, Boyi Tang, *Student Member, IEEE*, Shuai Gao, *Student Member, IEEE*, Kin-Fai Tong, *Fellow, IEEE*, Hang Wong, *Fellow, IEEE*, Kai-Kit Wong, *Fellow, IEEE* and Yangyang Zhang

*Abstract*—While multiple-input multiple-output (MIMO) technologies continue to advance, concerns arise as to how MIMO can remain scalable if more users are to be accommodated with an increasing number of antennas at the base station (BS) in the upcoming sixth generation (6G). Recently, the concept of fluid antenna system (FAS) has emerged, which promotes position flexibility to enable transmitter channel state information (CSI)-free spatial multiple access on one radio frequency (RF) chain. On the theoretical side, the fluid antenna multiple access (FAMA) approach offers a scalable alternative to massive MIMO spatial multiplexing. However, FAMA lacks experimental validation and the hardware implementation of FAS remains a mysterious approach. The aim of this paper is to provide a novel hardware design for FAS and evaluate the performance of FAMA using experimental data. Our FAS design is based on a dynamically reconfigurable "fluid" radiator which is capable of adjusting its position within a predefined space. One single-channel fluid antenna (SCFA) and one double-channel fluid antenna (DCFA) are designed, electromagnetically simulated, fabricated, and measured. The measured radiation patterns of prototypes are imported into channel and network models for evaluating their performance in FAMA. The experimental results demonstrate that in the 5G millimeter-wave (mmWave) bands (24-30 GHz), the FAS prototypes can vary their gain up to an averaged value of 11 dBi. In the case of 4-user FAMA, the double-channel FAS can significantly reduce outage probability by 57 % and increases the multiplexing gain to 2.27 when compared to a static omnidirectional antenna.

## I. INTRODUCTION

IN today's era of wireless communications, multi-input multi-output (MIMO) technology takes center stage, offering numerous advantages, including high diversity gain and low outage probability. In contrast, in traditional single input, single output (SISO) systems, deep fades in the signal can lead to communication failures due to a significant drop in the signal-to-noise ratio (SNR). Recently, fluid antenna systems (FAS) have emerged [1]–[4], where the fluid radiators can be dynamically positioned to different locations over a defined space. With the channel state information, the antenna's radiating element selects the position with strongest signal to communicate [5]–[7]. Studies on the channel model and analysis of FAS have demonstrated that the enhancements offered by FAS are significant [8]–[13]. As theoretically demonstrated in [14], a MIMO system in a space measuring $0.5\lambda$ by $0.5\lambda$ (where $\lambda$ denotes the wavelength) at both transmitter and receiver sides achieves a maximum diversity gain of 16. However, by combining MIMO with FAS in the same space, the diversity gain can be significantly increased to 81.

The complementarity of FAS with other wireless technologies opens up a fertile ground for innovation in future wireless systems, including 6G. Several MIMO-FAS scenarios have been reported [4], [14]–[17], in which FAS was considered at the user end. This integration enables MIMO systems to achieve exceptionally high data rates and reliability. FAS can also be synergistically combined with reconfigurable intelligent surfaces (RIS) and artificial intelligence [18], [19]. While the future of fluid antennas holds significant promise, the physical realization of a functional FAS remains a challenge.

Fluid antenna multiple access (FAMA) is a novel approach to handling interference [16], [20], [21]. It utilizes the radiator-position flexibility of fluid antennas to exploit natural fluctuations in wireless signals. By switching to a position where interfering signals fade, FAMA allows interference-less reception without complex signal processing. This could lead to simpler communication systems and potentially support a massive number of users sharing the same physical data channel. Despite the encouraging results of FAMA, the experimental validation is lacking.

Researchers have been exploiting the unique attributes of fluid, such as flexibility and adaptability, in antenna design. This approach utilizes either conductive or dielectric fluid material, allowing them to alter their shape and properties. Various approaches, including using liquids for reconfigurable switching and conductive connections, have been explored [22]–[27]. These approaches have demonstrated control over different aspects like frequency [28], [29], radiation pattern [30], [31] and polarization [32], [33]. However, fluid antennas equipped with the capability to select their radiator positions

This work was supported in part by the Engineering and Physical Sciences Research Council (EPSRC) under Grant EP/V052942/1 (Corresponding author: Kin-Fai Tong)

Yuanjun Shen, was with the Department of Electronic and Electrical Engineering, UCL, and now is with National Key Laboratory of Antennas and Microwave Technology, Department of Electronic Engineering, Xidian University, Xi'an, 710071, China (e-mail: yuanjun.shen@xidian.edu.cn).

Boyi Tang, Kin-Fai Tong, and Kai-Kit Wong, are with the Department of Electronic and Electrical Engineering, University College London, London, WC1E 7JE UK.

Shuai Gao and Hang Wong, are with the State Key Laboratory of Terahertz and Millimeter Waves, Department of Electrical Engineering, City University of Hong Kong, Hong Kong.

Yangyang Zhang is with KuangChi Science Limited, Hong Kong SAR, China.

for improving outage probability and multiplexing gain are yet to be reported.

This work bridges such a crucial gap in FAS research by introducing two novel surface wave enabled fluid antennas with a key innovation: dynamically positionable radiator elements. Unlike previously reported fluid antennas, these proposed designs offer the first physical realization of dynamic radiator positioning which enables the antennas to optimize their performance by selecting fluid radiator positions that minimize outage probability and maximize multiplexing gain. Furthermore, this work incorporates experimental results into channel and network models, allowing for rigorous verification of the performance improvements achieved by FAS in multi-user scenarios within the 5G millimeter-wave bands. Section II details the antenna geometry, including the surface wave launcher and 3D-printed fluid container housing the channel(s), along with electromagnetic simulation results. The operational principle and antenna matching are also discussed in this section. Section III presents the measured results of the two surface wave enabled fluid antenna prototypes. These measured results are then incorporated into the channel and network models for analysis. The improvement achieved in the proposed FAMA system is presented in Section IV. Finally, Section V concludes the work based on the obtained results.

## II. Antenna Design and Operation Principle

### A. Surface Wave Enabled Fluid Antenna Geometries

The perspective views of a single-channel fluid antenna (SCFA) and a double-channel fluid antenna (DCFA) are shown in Fig. 1. Each antenna consists of a PCB substrate (Rogers 5880, $\varepsilon_r = 2.2$, thickness = 0.787 mm and $\tan\delta = 0.0009$ at 10 GHz) with a conductor ground plane on the bottom side, a PCB surface wave launcher connected to a K connector, a fluid container with embedded fluid channel(s), and a fluid radiator inside each channel. The PCB surface wave launcher pattern shown in Fig. 2(d) is laser engraved on one end of the PCB substrate. The fluid container, which is 3D-printed by using the resin-based micro-stereo-lithography technology (epoxy resin, $\varepsilon_r = 4.0$, x-y-z printing resolution = 0.01 mm), contains the fluid channels to accommodate the fluid radiator motion. While Metallic or dielectric materials can be used as fluid radiators, fluid metal 'Galinstan'(electric conductivity = $3.46 \times 10^{-6}$ S/m, thermal conductivity = 16.5 W/(K m), material density = 6440 kg/m$^3$, thermal diffusivity = $8.655\,78 \times 10^{-6}$ m$^2$/s) is selected as the fluid radiator in this paper, as it has good scattering capability. The sizes of the antennas are small with overall dimensions of 33 × 10 × 2.8 mm$^3$ ($L_{substrate} \times W_{substrate} \times (H_{container} + H_{substrate})$). Fig. 2 illustrates the detailed structures including top and side views, together with the surface wave launcher pattern and assembled prototypes of the fluid antennas. The values of all parameters have been included in the caption. To enable the motion of the fluid radiator, air is pumped in or out through the inlets/outlets shown in Fig. 2(a) and Fig. 2(b) by a 3V peristaltic micro-pump (11.9 × 13.9 × 32.9 mm$^3$) [34] through thin silicon tubes. For high-precision control, the two ends can be connected to a fluid control system for lab-on-a-chip devices [35]. It is worth noting that pump-free electrowetting techniques can also be utilized to continuously traverse the fluid radiator at a speed up to 10 mm/s [36]. In the simulation model, the coplanar waveguide (CPW) feed line of surface wave launcher is extended to the edge of PCB for direct connection to a K connector as illustrated in Fig. 2(d). Therefore, the impact of the K connector on the radiation pattern is considered. The K connector can operate up to 40 GHz, enough to cover the target frequency range.

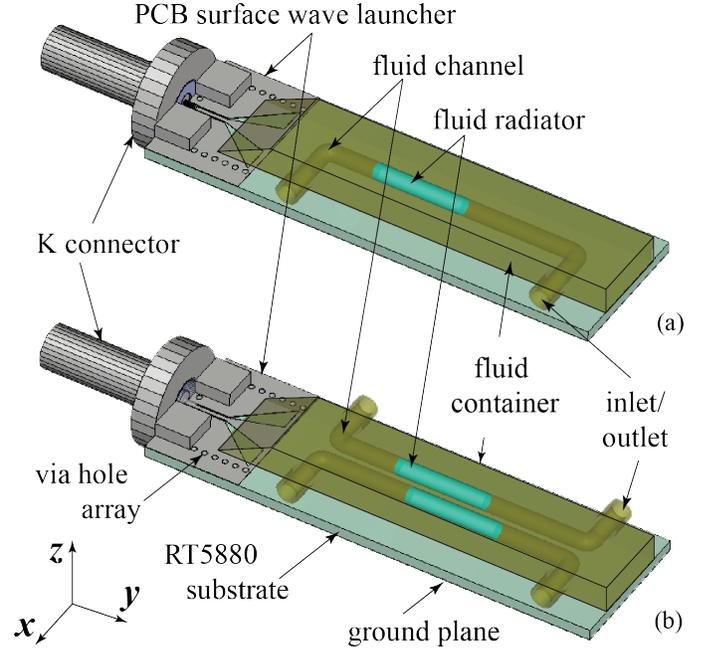

Fig. 1: The perspective views of (a) single-channel (SCFA), and (b) double-channels (DCFA), surface wave enabled fluid antennas.

### B. Surface Wave Launcher and 3D-printed Fluid Container

The PCB surface wave launcher converts the RF signal from the K connector (feeding port) to surface wave propagating on the substrate surface. To reduce the antennas' overall thickness, a planar launcher reported in [37] is simplified for exciting the surface wave in this paper, rather than using a wideband waveguide surface wave transducer [38]. The pattern and parameters of surface wave launcher are shown in Fig. 2(d). It is composed of a piece of PCB substrate with a copper ground plane on the bottom, a substrate-integrated waveguide (SIW) via hole array with width $W_{swl}$ and length $L_{swl}$ and a Y-shaped slot pattern.

The size of the 3D-printed fluid container is minimized to reduce its impact on the antenna performance. On the end contacting the surface wave launcher, a ramp with an optimized angle $\alpha_{con} = 26°$ is realized to provide a smooth transition avoiding undesirable impedance mismatch. An air channel of diameter $D_{channel} = 1.2$ mm and length $L_{channel} = 17.5$ mm is printed inside the container for the motion of the fluid radiator. As the length of the fluid radiators ($L_{rad}$, $L_{rad1}$ and $L_{rad2}$) is 6.5 mm, the possible continuous motion of the fluid radiator ($L_{shift}$) ranges from 0 to 11 mm, i.e., about 1 free space wavelength at 26 GHz.

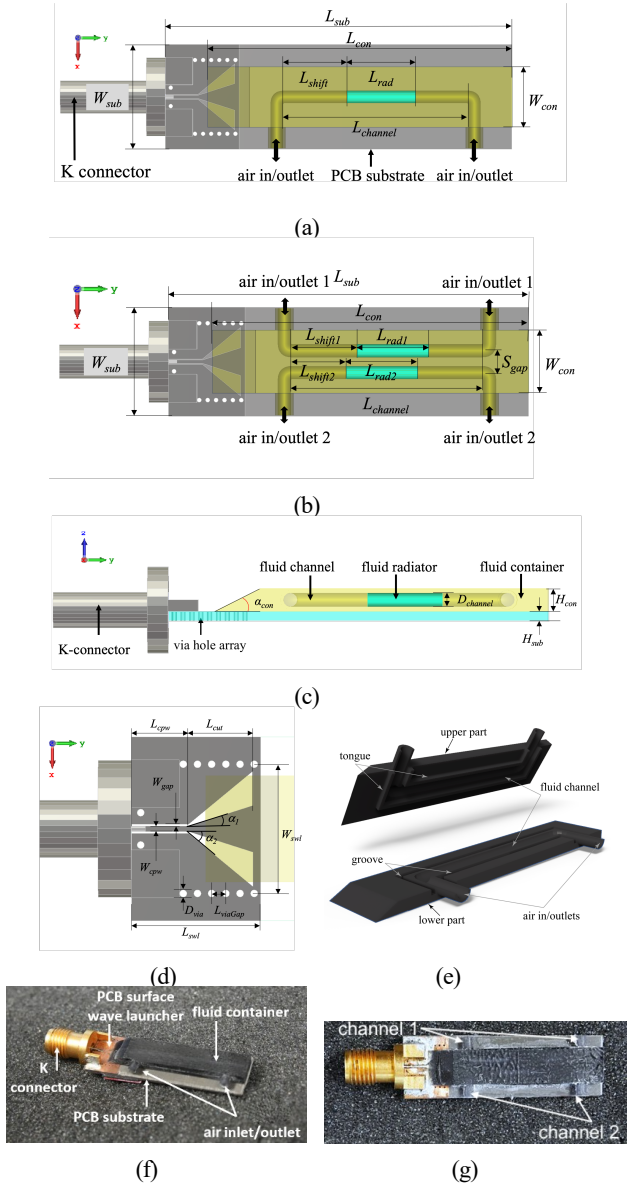

Fig. 2: The top view of (a) single-channel (SCFA) and (b) double-channel fluid antennas (DCFA), (c) side view, (d) surface wave launcher pattern, (e) opened view of the fluid container, (f) the assembled prototype of single-channel fluid antenna, and (g) the assembled prototype of double-channel fluid antenna. $L_{sub}$ = 33.0 mm, $L_{con}$ = 29.0 mm, $W_{sub}$ = 10.0 mm, $W_{con}$ = 5.8 mm, $H_{sub}$ = 0.787 mm, $H_{con}$ = 2.0 mm, $D_{channel}$ = 1.2 mm, $L_{channel}$ = 17.5 mm, $L_{rad}$ = $L_{rad1}$ = $L_{rad2}$ = 6.5 mm, $S_{gap}$ = 2 mm, $\alpha_1$ = 20°, $\alpha_2$ = 42°, $\alpha_{con}$ = 26°, $D_{via}$ = 0.4 mm, $L_{via}$ = 0.8 mm, $L_{cpw}$ = 1.5 mm, $W_{cpw}$ = 0.2 mm, $L_{swl}$ = 7.0 mm, $W_{swl}$ = 7.0 mm, $L_{cut}$ = 3.5 mm, $G_{cpw}$ = 0.08 mm.

The fluid containers are 3D-printed by a resin-based Nano-Dimension® DragonFly IV printer. However, printing hollow geometries without supporting structures always results in deformation. To maintain the precision and cylindrical shape of the hollow channels, the fluid containers are split into upper and lower halves at the midpoint of the channel as shown in Fig. 2(e) when it is printed. Potential leakage issue has also been considered. Groove and tongue pairs are implemented around the channels. They not only prevent any fluid leakage, but improve the alignment and increase the contact surface between the upper and lower parts. UV glue is used for assembling the fluid antennas allowing enough time for better alignments before curing the glue with UV light. Moreover, to maintain the smooth motion of Galinstan fluid radiator, a small amount (approximately 2 ml) of dilute aqueous sodium hydroxide is added to the channel around the Galinstan fluid radiator to keep it in a fluid state.

### C. Operation Principle

Fig. 3 shows the root-mean-square (RMS) value of E-field distributions of a single-channel fluid antenna (SCFA) at 26 GHz. The signal from the K connector is fed into the surface wave launcher which excites the signal into surface wave propagating along the positive y-direction on the surface wave platform. It can be observed that the E-field intensity is stronger at the area near the fluid radiator. In other words, the wave is scattered at the fluid radiator into free space as shown in Fig. 3(b) to 3(d). As the fluid radiator can be traversed along the fluid channel, the wave can be scattered at different positions, creating the required spatial diversity. The scattering of the wave by the fluid radiator also alters the radiation patterns. The E-field distribution of the model without the fluid radiator is provided in Fig. 3(a) for references.

Fig. 4 shows the 3D radiation patterns of selected representative fluid antenna models, including empty channel, i.e., without any fluid radiator, two SCFAs with fluid radiator positioned at the middle and far end of the fluid channel ($L_{shift}$ = 6 or 11 mm), and Case 1 ($L_{shift1}$ = 2 mm, $L_{shift2}$ = 8 mm) of DCFA. It can be observed from Fig. 4(a) to 4(c) that the radiation patterns are basically symmetric along the yz-plane for SCFA, while the symmetry disappears in Case 1 of DCFA. Such an asymmetric arrangement provides a higher degree of freedom for tackling different wireless channel conditions.

## III. ANTENNA MEASUREMENT RESULTS

Fig. 5(a) presents the comparison of the $S_{11}$ of SCFA with 12 distinct fluid radiator positions $L_{shift}$, ranging from 0 to 11.0 mm with 1.0 mm increments, while Fig. 5(b) shows the corresponding plot for the results of DCFA with 144 possible fluid radiator positions. It can be observed that the antennas' operational frequency ranges from 23.2 GHz to 38.5 GHz. However, the focus of this study is more in the millimeter-wave frequency range from 24.25 to 29.5 GHz, the Bands n257 and n258 for the 5G frequency band, so the antenna performance at 26 GHz is selected for demonstration. The highlighted thicker lines distinguish the representative cases. The $S_{11}$ of the antenna with empty channel(s) is depicted as the black line in Fig. 5 for reference. The thin dotted lines show the results of the remaining radiator position combinations. Echoing with the E-field distributions shown

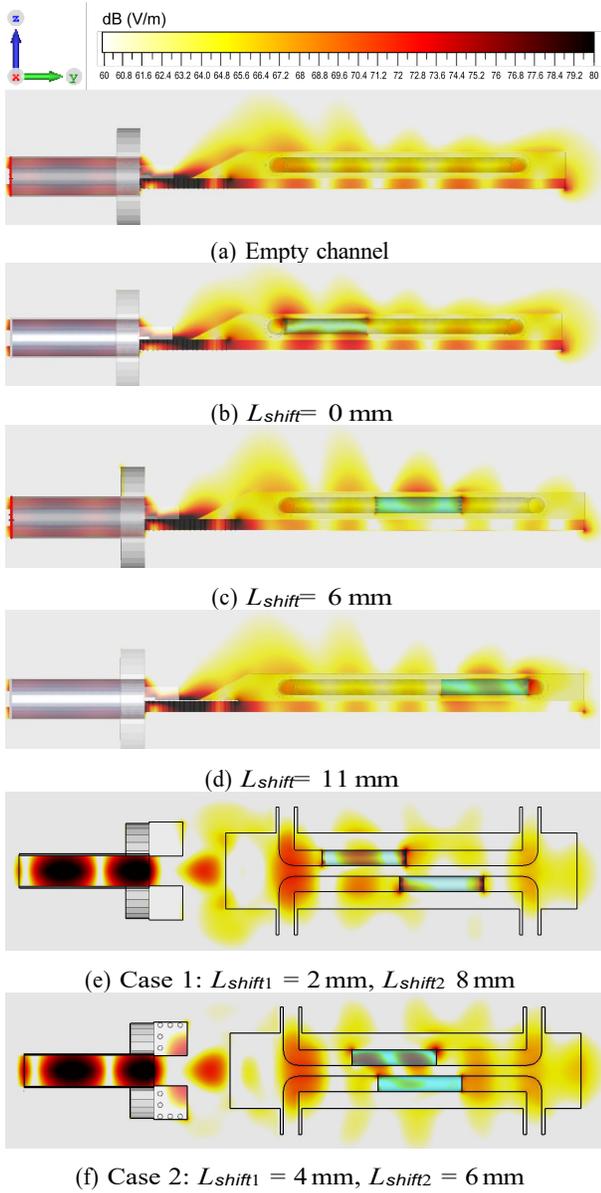

(a) Empty channel

(b) $L_{shift}$ = 0 mm

(c) $L_{shift}$ = 6 mm

(d) $L_{shift}$ = 11 mm

(e) Case 1: $L_{shift1}$ = 2 mm, $L_{shift2}$ 8 mm

(f) Case 2: $L_{shift1}$ = 4 mm, $L_{shift2}$ = 6 mm

Fig. 3: E-field distribution (RMS) of (a)-(d) SCFA and (e)-(f) DCFA, at different fluid positions at 26 GHz.

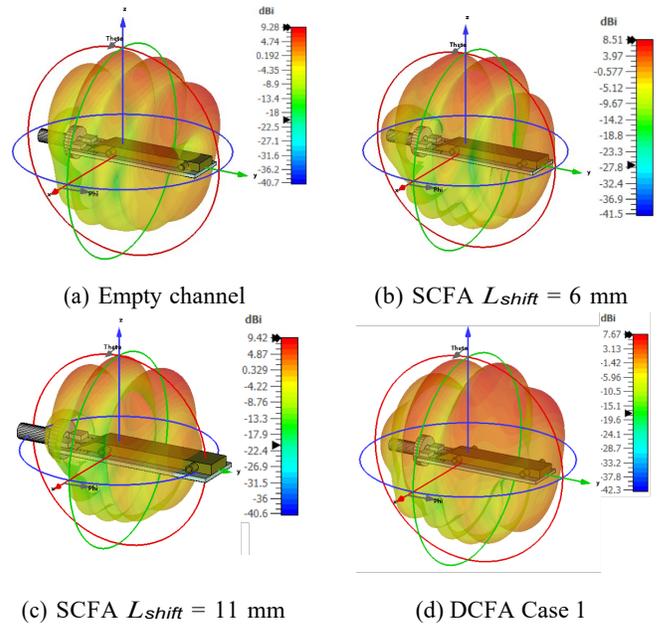

(a) Empty channel

(b) SCFA $L_{shift}$ = 6 mm

(c) SCFA $L_{shift}$ = 11 mm

(d) DCFA Case 1

Fig. 4: The 3D radiation patterns of the proposed fluid antenna with different fluid radiator positions.

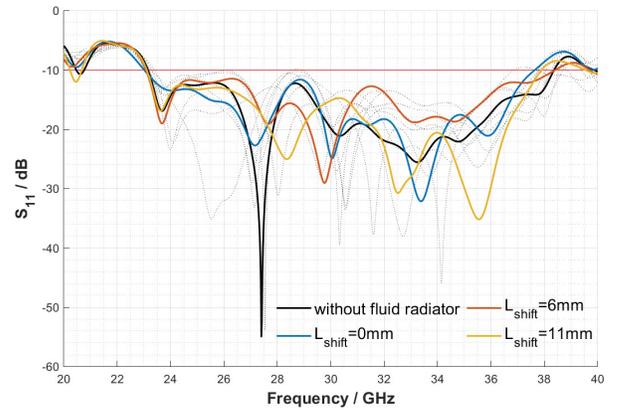

(a) SCFA

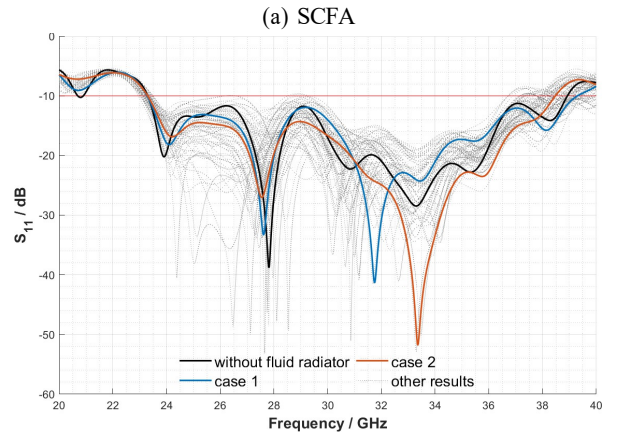

(b) DCFA

Fig. 5: The $S_{11}$ results of the proposed fluid antennas (a) single channel, and (b) double channels, with fluid radiators at different positions.

in Fig. 3, the realized gains and the corresponding radiation pattern of the antenna vary with the fluid radiator positions as shown in Fig. 6. In the case without the fluid radiator, the gain of radiation lobes gently increases in the angular range $(\theta, \phi)$ from $(-45°, 270°)$ to $(45°, 90°)$. A general pattern can be observed that the realized gain of the SCFA drops in the direction corresponding to the fluid radiator position due to stronger scattering at the fluid radiator. For example, at $(\theta, \phi) = (45°, 90°)$, the realized gain in the case without the fluid radiator is about 9.2 dBi, but it drops to 8.5 dBi when $L_{shift}$ is 0 mm. When $L_{shift}$ is increased to 6 mm, the realized gain further drops down to 5.0 dBi. However, it is important to distinguish this feature from beamsteering in phased array antennas. Here, the gain variation with fluid radiator position reflects pattern shaping, not a linear relationship for directing the main beam.

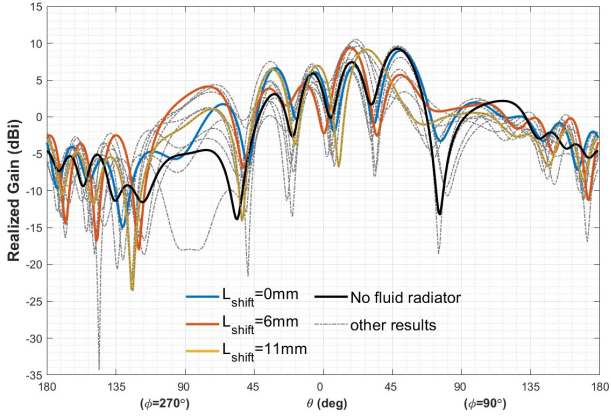

Fig. 6: Radiation pattern at the cutting plane of $\phi = 90°$ at 26 GHz of the SCFA at different fluid radiator positions.

To further explain the capabilities of the proposed fluid antennas, two scenarios are presented in Fig. 7. In the first scenario with two user equipment (UEs) and a base station. UE 1 needs to establish a reliable connection with the base station, but the signal for UE 2, located at $(\theta, \phi) = (20°, 270°)$ will also be received by UE 1 as shown in Fig. 7(a). Therefore, there will be interference between the two UEs. In this case, the fluid radiator on UE 1 can shift to $L_{shift} = 3$ mm in the second scenario. The corresponding result in Fig. 7(b) shows that the signal for UE 2 is significantly reduced to a null. Thus, the fluid antenna can minimize interference caused by UE 2. It is important to note that this scenario represents only a specific case with notable improvements. On the other hand, when a UE approaches the null direction of a fluid antenna, the fluid radiator should shift to a different position, preventing the UE from experiencing a prolonged null state. These two features are particularly favorable in mobile communications. To the authors' best knowledge, there are currently no single antenna elements that can perform such adaptive capability to effectively address such scenarios. A more comprehensive study, which includes 1 million randomly generated wireless channel realization models, will be provided in Section IV to show the fluid antennas' unique capabilities.

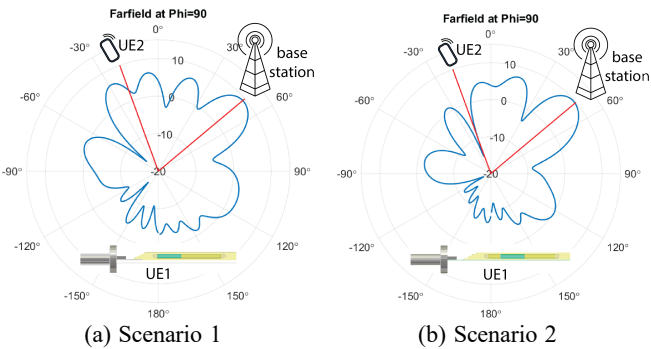

Fig. 7: The radiation pattern comparison of the antenna with the fluid radiator at (a) $L_{shift} = 0$ mm and $L_{shift} = 3$ mm at 26 GHz at the cutting plane of $\phi = 90°$

The maximum and minimum values of the antenna gain across different fluid radiator positions on the yz-plane are superimposed as the red (upper envelop) and black (lower envelop) curves as shown in Fig. 8(a) and 8(b). These upper and lower envelops are combined into a unified figure, facilitating the determination of the approximate range of achievable gain variation as the radiator traverses the fluid channel. Here, we define the possible gain variation between the upper and lower envelops as the radiation pattern dynamic range (RPDR). It can be observed that the RPDR of the SCFA is narrower than that of DCFA, reflecting the higher degree of freedom in DCFA. The averaged RPDR at different fluid radiator positions on the yz-plane in the target frequency band are plotted in Fig. 8(c). It can be observed that the antennas exhibit a maximum average RPDR of up to 11 dBi.

## IV. FAMA

To quantify the strength of the proposed fluid antennas in improving outage probability and multiplexing gain in future mobile communications, a network model and a channel model have been developed to analyze the performance of SCFA and DCFA in multi-user mobile communication environments. In the models, we consider optimization problems in a downlink FAMA [39]–[43] arrangement. There are one base station (BS), and $M$ users ($M \leq 4$) in the system. Each UE is assumed to have the proposed antenna whose fluid radiator element can be traversed to one of the $N$ evenly separated preset locations over a linear dimension of length $W\lambda$, where $W$ denotes the normalized length of fluid antenna and $\lambda$ is the wavelength. The preset locations are referred to as ports and the radiating element at each port is treated as the repositionable fluid radiator in either a SCFA or DCFA. The BS is equipped with $M$ fixed antennas and each antenna of the BS serves one users. Without loss of generality, we assume the information symbol dedicated for the $i^{th}$ user is transmitted by the $i^{th}$ antenna of BS. Therefore, the signal from the $j^{th}$ ($j \neq i$) antenna of BS is considered as interference at the $i^{th}$ user. The signal to interference plus noise ratio (SINR) of the $i^{th}$ UE is given by

$$SINR_i = \frac{p_i |h_{i,i}^{(k_i^*)}|^2}{\sum_{j \neq i}^{M} p_j |h_{j,i}^{(k_i^*)}|^2 + \sigma_\eta^2}, \quad (1)$$

where $k_i^*$ denotes the selected port of fluid antenna of the $i^{th}$ UE.

To model the signal from different angle-of-arrival (AoA), we use the finite scattering channel model in [40]. The channel is modelled with a specular component, i.e., line-of-sight (LoS) and $N_p$ scattered components (i.e., non-LoS)

$$g_{j,i}^{(k)} = \sqrt{\frac{K\Omega}{K+1}} e^{j\alpha_{j,i}} e^{-j\frac{2\pi(k-1)W}{N-1}\cos\theta_{j,i}^0} + \sum_{l=1}^{N_p} \alpha_{j,i}^l e^{-j\frac{2\pi(k-1)W}{N-1}\cos\theta_{j,i}^l}, \quad (2)$$

where $g_{j,i}^{(k)}$ denotes the theoretical channel from the $j^{th}$ antenna of BS to the $k^{th}$ port of the $i^{th}$ UE, $K$ is the power ratio between the specular and scattered components, $\alpha_{j,i}^l$ is the random phase of the specular component, $\alpha_{j,i}$ is the random complex coefficient of the $l^{th}$ scattered path, $\theta_{j,i}^0$ and $\theta_{j,i}^l$ denotes the AoA of the LoS and the $l^{th}$ non-LoS from the $j^{th}$ antenna of BS to the $i^{th}$ UE respectively, $E[|g_{j,i}^{(k)}|^2] = \Omega$ and $E[\sum_l |\alpha_{j,i}^l|^2] = \frac{\Omega}{K+1}$. Consider the radiation pattern of fluid antenna, the received channel at the UE is:

$$h_{j,i}^{(k)} = \sqrt{\frac{K\Omega}{K+1}} e^{j\alpha_{j,i}} e^{-j\frac{2\pi(k-1)W}{N-1}\cos\theta_{j,i}^0} \overline{Gain_{\theta_{j,i}^0}^k} + \sum_{l=1}^{N_p} \alpha_{j,i}^l e^{-j\frac{2\pi(k-1)W}{N-1}\cos\theta_{j,i}^l} \overline{Gain_{\theta_{j,i}^l}^k}, \quad (3)$$

where $Gain_\theta^k$ denotes the antenna gain in the direction of $\theta$ at the $k^{th}$ port of fluid antenna.

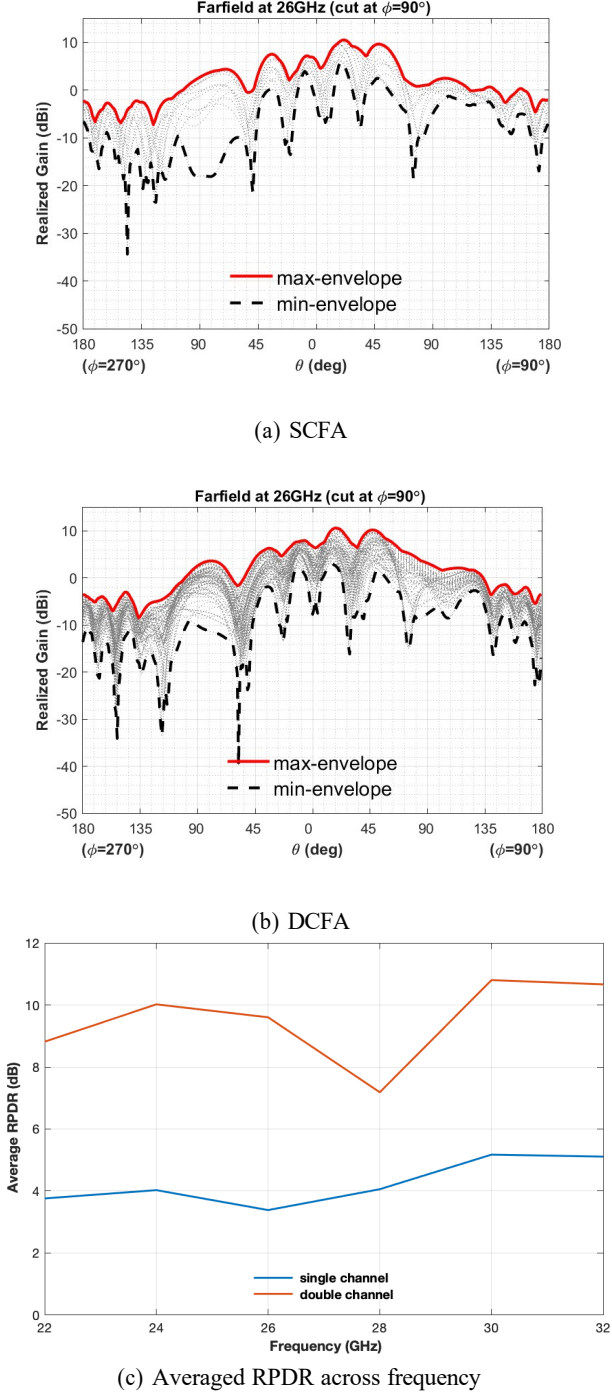

(a) SCFA

(b) DCFA

(c) Averaged RPDR across frequency

Fig. 8: Radiation pattern dynamic ranges of (a) SCFA, (b) DCFA, and (c) averaged RPDR against frequency.

This section presents the outage possibility and multiplexing gain, as defined in [40], of the proposed SCFA and DCFA. These fluid antennas operate in a multi-user mobile communication environment with $K=20$, $\Omega=1$, and $N_p=5$. In the case of SCFA, 20 radiation patterns were measured at different fluid radiator positions, corresponding to 20 positions separated by 0.5 mm increments with $L_{shift}$ ranges from 0 to 9.5 mm in an anechoic chamber. For the DCFA, there are 12 fluid radiator positions separated by 1 mm increments in each channel ($L_{shift1}$ and $L_{shift2}$ range from 0 to 11 mm). Therefore, there are 144 possible combinations of fluid radiator positions. The measured radiation patterns of the two fluid antennas at 26 GHz are imported into the theoretical models. The results for ideal omni-directional (OD) antennas are also simulated in the models for comparison. One million randomly generated channel realization and port position simulations are performed in each case. The results aim to provide two important insights. First, the results can be considered as the generalized performance of common static antenna position approach if the antenna is located at one of the many positions. Meanwhile, the unique feature of fluid antenna can be identified by comparing the results with the static approach. This is because, unlike static antennas, the fluid radiator position(s) can strategically shift based on the channel condition, making the antenna operation dynamic. Different numbers of UEs, from 1 to 4, experiencing various SINR levels are modeled in the simulations.

As in [42], we define an outage event if the SINR of user is below the target $\gamma$. Then the outage probability is defined as $Prob(SINR < \gamma)$. In our simulation, we consider $\gamma = 0$ dB, which means that the power of desired signal is larger than the power of interference signal and noise. The multiplexing gain of the system is defined as the capacity scaling factor, which is given by

$$m = M(1 - Prob(SINR < \gamma)), \quad (4)$$

where $M$ is the number of UEs in the system.

Fig. 9 shows the outage possibility of the multi-user system when SCFA and DCFA are used. In the single UE case, omni-directional antennas have lower outage probability compared to that of SCFA/DCFA since the gain of SCFA/DCFA is lower than 0 dBi across 90% of their angular coverage as shown in Fig. 6. This means the fluid antennas receive less desired signal in most direction. It is worth noting that the outage possibility of both dynamic cases (omni-directional and fluid antennas) outperform that of the static cases in multi-user cases ($M > 1$). With an increasing number of UEs, the omni-directional antennas experiences higher outage probability than the fluid antennas. This is because the power of desired signal and the interference are all influenced by the gain of antennas. In particular, for SCFA/DCFA, the gap between the desired signal and interference may become larger than with omni-directional antenna, as depicted in the RPDR. Additionally, the outage possibility may also have higher value due to the increased number of UE. In the case of 4 UEs, the outage possibility of the system using dynamically positionable DCFA drops to 0.43, which means it is 57% better than the 1.0 of the omni-directional static case. For the single user case, the UE can always reach the SINR requirement as the transmit SNR increases. However, as the number of UEs increases, the UE receives more interference, leading to a decrease in SINR and higher frequency of the outage events. For the 2 UEs case, the outage probability of generalized fixed port selections is around 0.5 for both antennas. This is improved to 0.15 (SCFA) and 0.056 (DCFA) with desired port selection. While the outage probability increases with the number of UEs, as expected, both SCFA and DCFA dynamic cases show a lower outage possibility compared to the static and omni-directional cases.

Fig. 10 shows the multiplexing gain. The single user system reaches the upper bound of 1 when the SNR (SINR equals SNR as no interference in the single UE case) approaches 15 dB. For the fluid antennas, the 2 or 3 UE cases show the best performance in multiplexing gain (1.7 for SCFA and 2.27 for DCFA). As the number of users increases, there is more interference in the system. The multiplexing gain of 4 UEs in SCFA reduces, likely due to the increase in outage probability for scenarios with more users. However, it is overcome by DCFA because of the higher dimension of diversity and higher PRDR. Overall, the multiplexing gain is enhanced by selecting the optimized fluid radiator position with maximum SINR compared to random selection.

## V. CONCLUSION

In this paper, we proposed two fluid antennas enabled by surface wave technology for future 6G mobile communication applications. The performance of the SCFA and DCFA were simulated and measured in the 5G millimeter-wave frequency bands. Due to the fluid nature of the radiators, their position can traverse the straight fluid channels. This unique feature allows for dynamic control of the antennas' radiating position and pattern, enhancing connectivity in different wireless environments. A new indicator RPDR is defined to describe the difference between the maximum and minimum gain of the fluid antenna in particular directions when the fluid radiators traverse the fluid channels. According to the results calculated from the channel and network models, the outage probability can be significantly reduced compared to the usual static antenna position approach in multi-user scenarios. Furthermore, the multiplexing gain can reach 1.7 and 2.27 in the cases of SCFA and DCFA, respectively.

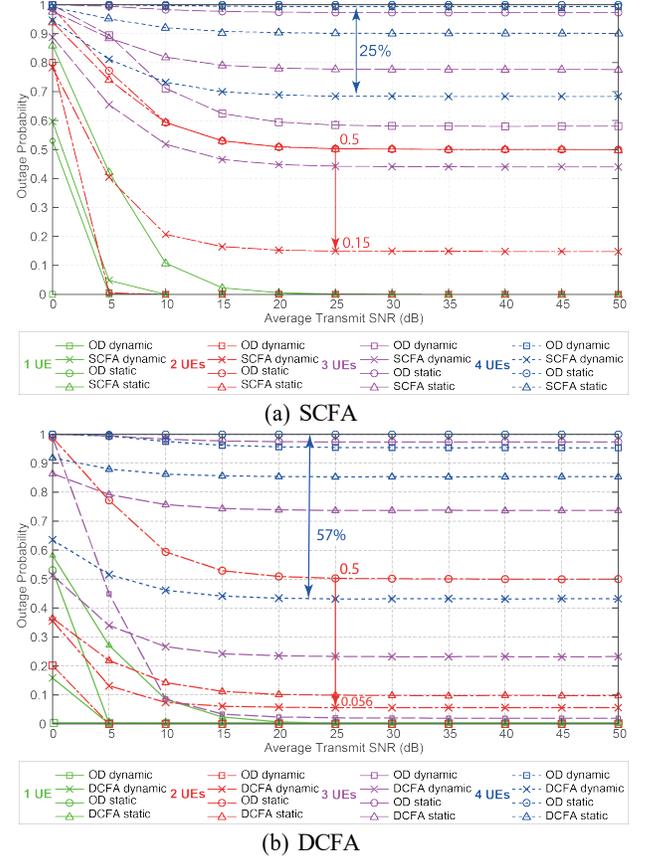

Fig. 9: Outage possibility of (a) SCFA, (b) DCFA.

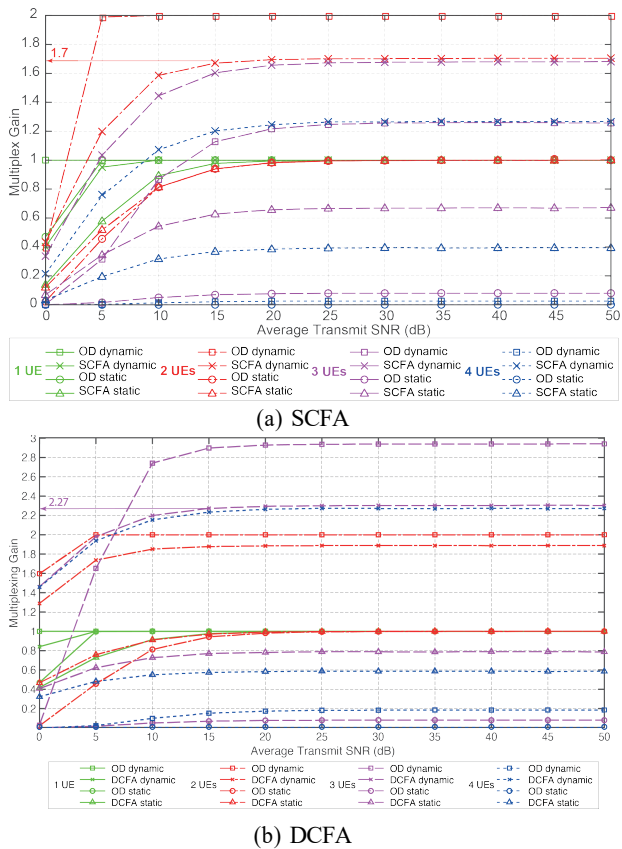

Fig. 10: multiplexing gain of (a) SCFA, (b) DCFA.

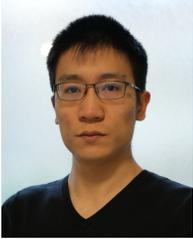

**Yuanjun Shen** (Member, IEEE) was born in Xi'an, China. He received the B.Eng. degree in electronic engineering from the Tianjin University, China, in 2014, the M.S. degree in electronic engineering from King's College London, UK, in 2016; and the Ph.D. degree in electronic and electrical engineering from University College London (UCL), UK, in 2023. From 2017 to 2023, he was a Research Assistant at UCL, London, UK. After graduation, he joined the National Key Laboratory of Antennas and Microwave Technology, Department of Electronic Engineering of Xidian University in 2023. His current research interests include millimeter-wave antennas, fluid antennas, gap waveguides, radar systems, and communications systems at mmW band.

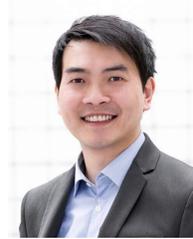

**Kai-Kit Wong** (M'01-SM'08-F'16) received the BEng, the MPhil, and the PhD degrees, all in Electrical and Electronic Engineering, from the Hong Kong University of Science and Technology, Hong Kong, in 1996, 1998, and 2001, respectively. After graduation, he took up academic and research positions at the University of Hong Kong, Lucent Technologies, Bell-Labs, Holmdel, the Smart Antennas Research Group of Stanford University, and the University of Hull, UK. He is Chair in Wireless Communications at the Department of Electronic and Electrical Engineering, University College London, UK. His current research centers around 6G and beyond mobile communications. He is Fellow of IEEE and IET. He served as the Editor-in-Chief for IEEE Wireless Communications Letters between 2020 and 2023.

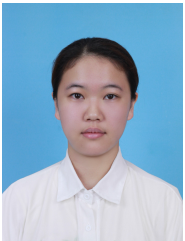

**Boyi Tang** (Student Member, IEEE) received the B.Eng. degree in communication engineering from the University of Electronic Science and Technology of China and the B.Eng. degree (Hons.) in electronics and electrical engineering from the University of Glasgow in 2022. She is currently pursuing the Ph.D. degree with the Department of Electronic and Electrical Engineering, University College London, U.K.. Her research interests include wireless communication, fluid antenna system, and mathematical optimization.

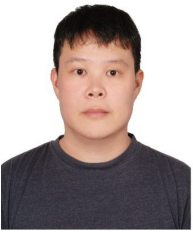

**Shuai Gao** (Graduate Student Member, IEEE) received the B.S. and M.S. degrees in electronic science and technology from Shenzhen University, Shenzhen, China, in 2017 and 2020, respectively. He is currently pursuing the Ph.D. degree in electrical engineering with the City University of Hong Kong, Hong Kong. His current research interests include transmitarrays, reflectarrays, shared-aperture antennas, and 6G antenna designs.

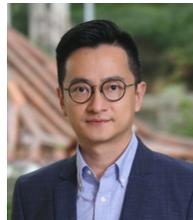

**Hang Wong** (Fellow, IEEE) received the B.Eng., M.Phil., and Ph.D. degrees in electronic engineering from City University of Hong Kong in 1999, 2002 and 2006, respectively. He joined the Department of Electrical Engineering at City University of Hong Kong in 2012. He is the director of Applied Electromagnetics Laboratory at CityU; and the deputy director of the State Key Laboratory of Terahertz and Millimeter Waves (Hong Kong). His research interests are antenna technologies of 5G, 6G, millimeter-wave and terahertz applications. He has over 270 publications, 2 co-authors of book chapters and 30 US and China patents. He was the chair of the IEEE Hong Kong Section of the Antennas and Propagation (AP)/Microwave Theory and Techniques (MTT) Chapter in 2011-2014 and 2021-2023. He was the IEEE APS Region-10 Representative from 2015-2021. He is an associate editor of IEEE Transactions on Antennas and Propagation and IEEE Antennas and Wireless Propagation Letters. Dr. Wong was the General Co-chair of the Asia Pacific Microwave Conference (AMPC) 2020, Hong Kong; the General Chair of Cross-Strait Radio Science and Wireless Technology Conference 2021, Shenzhen, China; the General Chair of 2025 IEEE International Workshop on Electromagnetics (iWEM), Hong Kong.

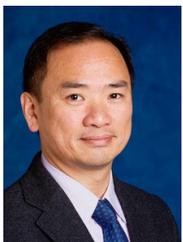

**(Kenneth) Kin-Fai Tong** (Fellow, IEEE) receive the B.Eng. and Ph.D. degrees in electronic engineering from the City University of Hong Kong. After graduation, he was with the Department of Electronic Engineering, City University of Hong Kong, as a Research Fellow. Two years later, he was an Expert Researcher with the Photonic Information Technology Group and the Millimeter-wave Devices Group, National Institute of Information and Communications Technology (NiCT), Japan, where his main research focused on photonic-millimeter-wave planar antennas at 10 GHz, 38 GHz, and 60 GHz, for high-speed wireless communications systems. In 2005, he was a Lecturer with the Department of Electronic and Electrical Engineering, UCL, where he is currently the Chair of Antennas and Applied Electromagnetics at University College London. His current research interests include millimeter-wave and THz antennas, fluid antennas, surface wave propagation, 3D printed antennas, and sub-GHz long range IoT networks. He served as the General Chair for the 2017 International Workshop on Electromagnetics (iWEM) London. He was a Lead Guest Editor of IEEE Open Journal of Antennas and Propagation in 2020.